\documentclass[
]{ceurart}

\usepackage{listings}
\usepackage{dsfont}
\usepackage{enumitem}
\usepackage{braket}
\usepackage{tabularx}
\numberwithin{equation}{section}

\begin{document}

\copyrightyear{2022}
\copyrightclause{Copyright for this paper by its authors.
  Use permitted under Creative Commons License Attribution 4.0
  International (CC BY 4.0).}

\conference{AIQxQIA 2026: International Workshop on AI for Quantum and Quantum for AI | co-located with AIxIA 2026, Perugia, Italy}

\title{Steering the Fidelity in State Preparation on Noisy Bosonic Systems with Risk-Seeking Optimization}

\author[1]{Riccardo Piantoni}[
email=riccardo.piantoni@mail.polimi.it,
]\cormark[1]
\author[1]{Riccardo Pellini}[
orcid=0009-0005-0435-0264,
email=riccardo.pellini@polimi.it,
]\cormark[1]
\author[1]{Maurizio {Ferrari Dacrema}}[%
orcid=0000-0001-7103-2788,
email=maurizio.ferrari@polimi.it,
]\cormark[1]
\author[1]{Paolo Cremonesi}[%
orcid=0000-0002-1253-8081,
email=paolo.cremonesi@polimi.it,
]\cormark[1]
\address[1]{Politecnico di Milano, Italy}

\cortext[1]{Corresponding author.}

\begin{abstract}
    In order to execute quantum algorithms on bosonic quantum computers, it is necessary to encode their inputs into a desired target quantum state, with a process called state preparation.  
    In real settings, state preparation methods operate in stochastic scenarios, where the prepared state is represented as a mixed state, that is a probabilistic ensemble of pure states.
    Therefore, existing state preparation methods optimize a variational ansatz by averaging over all possible sequences of photon loss events contained in the prepared ensemble, yielding an output state which approximates the desired target with a certain average fidelity. Such an approach, however, does not take into account directly the distribution of fidelities across states in the ensemble.
    In this work, we aim to maximize the fidelity of only a subset of pure states sampled from the prepared ensemble, at the cost of reducing the average fidelity overall. 
    This trade-off is relevant when the goal is to obtain high-fidelity states with a certain probability, rather than maximizing the average fidelity of the prepared ensemble.
    We borrow and adapt objective functions from risk-sensitive optimization, a framework that optimizes specific characteristics of a distribution rather than only its expectation, and that has been successfully applied in Reinforcement Learning.
    In the considered experimental settings, our results show that the infidelities of some pure states in the prepared ensemble can be reduced by up to four orders of magnitude when using risk-seeking objectives, compared to those achieved using the standard optimization of the average fidelity. Furthermore, we show that it is possible to control the trade-off between the probability of obtaining high-fidelity states and the lowest infidelity threshold reached, providing the flexibility to select the operating point that best suits a certain application.
\end{abstract}

\begin{keywords}
    Quantum Computing \sep
    Bosonic Systems \sep
    State Preparation \sep
    Reinforcement Learning \sep
    Risk-Sensitive Optimization
\end{keywords}

\maketitle

\section{Introduction}

Quantum computing has been attracting a lot of interest for its potential to provide speed-ups in tackling specific classes of problems hard to solve with classical computers. While qubits extend classical bits to the quantum domain, qudits further generalize them to multi-level quantum systems, exploiting the intrinsic non-binary nature of quantum mechanics.
In this context, several quantum algorithms require that the state of the input qudits is initialized to their input data, with a process called state preparation.
In bosonic quantum computers, a particular family of devices which work on qudits, the SNAP-Displacement (SD) protocol \cite{heeres_cavity_2015, krastanov_universal_2015, fosel_efficient_2020, ogunkoya_investigating_2023}, allows one to prepare arbitrary target states through variational optimization of a parametric ansatz, named SD-ansatz.
When working in noisy and realistic conditions, however, state preparation must take into account the stochasticity introduced by noise. In bosonic systems, the dominant source of noise is the photon loss that occurs inside a physical cavity, which is modeled using one of multiple equivalent \emph{Kraus representations} \cite{kraus_first_1983, nielsen_quantum_2010}. 
Consequently, the evolution of a quantum state in noisy conditions is non-deterministic and is not described by a single pure state, but rather by a \emph{mixed state}, defined as a probabilistic ensemble of outcomes, each associated to a specific sequence of photon loss events. Such a mixed state is described by a density matrix, that can be expressed as the expectation over all the possible sequences of photon loss events, called \emph{quantum trajectories} \cite{ciccarello_quantum_2022, wiseman_quantum_1996, lidar_lecture_2020, dalibard_wave-function_1992, dum_monte_1992, bompais_parameter_2022, castin_wave_2008, hamazaki_introduction_2026}, each one representing a pure state.
In the presence of noise, the typical way to optimize the gate parameters of ansatzes, such as those of the SD protocol, is to maximize the fidelity 
between this output density matrix and the target state, rather than between two pure states \cite{kudra_robust_2022, dacrema_sparse_2026}. Given the definition of the density matrix, such an objective is equivalent to maximizing the expected fidelity across all quantum trajectories, which we refer to as \emph{average fidelity}.
This means that the parameters of the SD-ansatz are optimized 
to maximize, on average, the fidelity over all possible trajectories.
As a result, multiple trajectories are aggregated into a single expectation value, leaving no direct control on the distribution of the fidelity across the individual noise trajectories.
While the overparametrization of the SD protocol \cite{fosel_efficient_2020} can be exploited to address this issue partially, there is still a degradation in the average fidelity
compared to the noiseless case \cite{dacrema_sparse_2026}, which strongly points to the presence of some intrinsic interference structure between noise trajectories. 
This raises the question of whether the optimization objective can instead be designed to shape the fidelity distribution over quantum trajectories, rather than only its expectation. Our goal is to increase the probability of generating high-fidelity quantum trajectories, even at the expense of the average fidelity of the density matrix of the output mixed state. In Reinforcement Learning (RL) \cite{howard_risk-sensitive_1972, heger_consideration_1994, mihatsch_risk-sensitive_2002, noorani_risk-sensitive_2021, marthe_efficient_2025}, Finance \cite{pratt_risk_1964, artzner_coherent_1999, follmer_convex_2002} and Optimal Control \cite{jacobson_optimal_1973, james_risk-sensitive_1994, dupuis_robust_2000, james_risk-sensitive_2004, abdelhafez_gradient-based_2019, azizpour_optimal_2025}, such a trade-off is addressed through risk-sensitive optimization \cite{pratt_risk_1964}, a family of methods that optimize specific properties of a reward distribution rather than only its expectation. In this work, we adapt this framework to noisy quantum state preparation by adopting a risk-seeking optimization strategy. 
More specifically, we explicitly account for sequences of photon loss events by sampling photon-counting quantum trajectories, according to the probabilities defined by the corresponding Kraus operators.
This allows us to employ trajectory-wise objective functions that directly shape the fidelity distribution. In particular, we investigate the application of the entropic risk measure \cite{howard_risk-sensitive_1972, marthe_efficient_2025} to the SNAP-Displacement protocol. 
Our results show that this risk-seeking formulation enables higher-fidelity state preparation while giving a controllable trade-off between reaching a target infidelity threshold and the probability of doing so.

\section{Background}
In this section, we briefly describe how the SNAP-Displacement (SD) protocol works, the structure of the SD-ansatz, and how the noise model is applied in this context. Then, we discuss the risk-sensitive optimization framework, and we touch upon techniques for gradient estimation in stochastic settings. 

\subsection{SNAP-Displacement Protocol}
\label{sec:snap_displacement}
Qudit states are vectors that live in a multidimensional Hilbert space $\mathcal{H}$, called \emph{Fock space}. The basis of this space is the set of orthogonal quantum states that physically correspond to any positive integer number $k$ of photons inside a cavity. We call such states \emph{Fock states} and denote them as $\ket{k}$, for $k \in \mathbb{N}$. A pure qudit state $\ket{\psi}$ is then an $\ell_2$-normalized superposition of Fock states. In bosonic quantum computing, universal control over qudit states \cite{krastanov_universal_2015} can be achieved by the combined action of Displacement operators $D(\alpha)$ and Selective Number-dependent Arbitrary Phase (SNAP) operators $S(\vec{\theta})$ \cite{heeres_cavity_2015, krastanov_universal_2015}.

\emph{Displacements} apply a shift to the amplitudes of the qudit state, controlled by the complex parameter $\alpha$, generally resulting in a new superposition of Fock states.
Mathematically, a Displacement gate is defined as: 
\begin{equation}
    D(\alpha) = e^{\alpha {a}^\dagger - \alpha^* {a}},
\end{equation}
where $\alpha \in \mathbb{C}$ is the displacement parameter and ${a}, {a}^\dagger$ are the annihilation and creation quantum operators respectively.
A \emph{SNAP} operation, instead, represents the simultaneous application of multiple phases $\vec{\theta} = [\theta_0, \theta_1, \theta_2, \cdots]$, with $\theta_k \in [0, 2\pi)$ for all $k \in \mathbb{N}$, one for each Fock state in the Fock basis, to the superposition of a qudit state. Mathematically, a SNAP gate is defined as
\begin{equation}
    S(\vec{\theta}) = \sum_{k=0}^{\infty} e^{i \theta_k} \ket{k}\bra{k}.
\end{equation}

In the SD protocol \cite{fosel_efficient_2020}, a parametric SD-ansatz is built as an alternating sequence of Displacement gates $D(\alpha)$ and SNAP gates $S(\vec{\theta})$. The $B$ pairs of consecutive gates form the \emph{SD-blocks} of the ansatz, each with its own time duration $T_b$.
In a noiseless scenario, the whole SD-ansatz of $B$ SD-blocks can be written as in Equation~\eqref{eq:snap_displacement_protocol}, where $\vec{\alpha}$ and $\Theta$ are sets of parameters optimized with variational strategies:
\begin{equation}
\label{eq:snap_displacement_protocol}
    U(\vec{\alpha}, \Theta) = D(\alpha_B) \prod_{b=0}^{B-1}S(\vec{\theta}_b)D(\alpha_b).
\end{equation}
In practice, current bosonic quantum devices can control only a finite number $d$ of Fock states, corresponding to photon numbers from $0$ to $d-1$. We therefore restrict the controllable space to ${\ket{0},\ldots,\ket{d-1}}$, referring to transitions to Fock states outside this space as \emph{leakage}. In our simulations, we include additional non-controllable Fock states, referred to as \emph{bumper states}, which allow us to account for part of this leakage.

\subsection{Noise Model}
\label{sec:noise_model}
In noisy settings, the effect of quantum noise can be modeled as a quantum channel, represented in terms of a set of \emph{Kraus operators} \cite{kraus_first_1983, nielsen_quantum_2010} $\{K_j,\,  j \in \mathbb{N}\}$. If the quantum state is represented by its density matrix $\rho$, its evolution under the noise channel is obtained by applying all the Kraus operators to $\rho$ and summing over the corresponding outcomes. The resulting density matrix describes a mixed state that accounts for all possible noise realizations. 
Formally, the evolution of $\rho$ due to such a quantum channel $\mathcal{E}$ is given by:
\begin{equation}
\label{eq:kraus_representation}
    \mathcal{E}(\rho) = \sum_{j \in \mathbb{N}} K_j \rho K_j^\dagger.
\end{equation}
There are infinite equivalent representations of Kraus operators that can be used to model a quantum channel \cite{kraus_first_1983, wiseman_quantum_1996}. In our setting, we model photon loss using its \emph{photon-counting Kraus representation}, where each Kraus operator $K_j$ directly corresponds to a specified amount $j$ of photons lost inside a physical cavity. As described for example in \cite{dacrema_sparse_2026}, photon-counting Kraus operators are defined as:
\begin{equation}
\label{eq:photon_loss_kraus_operators}
    {K}_j = \sqrt{\frac{(1-\eta)^j}{j!}} \eta^{{n}/2} {a}^j, 
\end{equation}
in which $\eta = e^{-\frac{t}{T_1}}$ is the survival probability of every single photon after an elapsed time equal to $t$ for a cavity having decoherence time $T_1$, and ${n} = {a}^\dagger {a}$ is the particle number operator that counts how many particles are present in the system.
The strength of the photon-loss channel is determined by the cavity decoherence time $T_1$: smaller values of $T_1$ correspond to stronger photon-loss effects. Since we apply the channel after every SD-block, the elapsed time $t$ amounts to the total duration $T_{b}$ of the SD-block.
However, representing the state of the qudit with a density matrix hides the internal probability distribution of the individual noise trajectories, which we instead aim to leverage. For this reason, we use an equivalent formulation \cite{ciccarello_quantum_2022}: we sample individual noise trajectories, in which the qudit state evolves in time as a pure state. This can be done by sampling and applying one Kraus operator to the pure state being propagated after each SD-block. Suppose that after block $b$ the quantum state is the pure state $\ket{\psi_b}$: then, the stochasticity of noise can also be modeled as the random application of one Kraus operator $K_j$ to $\ket{\psi_b}$. In particular, the probability of sampling the Kraus operator $K_j$ is given by
\begin{equation}
\label{eq:kraus_sampling}
    p_j\left(\ket{\psi_b}\right) = \left| \left | K_j \ket{\psi_b} \right | \right | ^ 2 \quad \forall j \in \mathbb{N},
\end{equation}
which, in our case, corresponds to the probability of losing exactly $j$ photons. Applying the Kraus operator $K_j$ results in an unnormalized pure state; therefore, we normalize it to get the sampled state following the photon loss event:
\begin{equation}
\label{eq:next_state_kraus_sampling}
    \ket{\psi_b}' = \frac{K_j \ket{\psi_b}}{\left | \left | K_j \ket{\psi_b} \right | \right |}.
\end{equation}
We define a \emph{quantum trajectory} $\tau$ as the sequence of pure states obtained by applying the sequence of SD-blocks, each followed by a single photon loss event, modeled by the corresponding Kraus operator. 
Specifically, starting from the initial state, after each SD-block a Kraus operator is sampled according to the probability distribution arising in Equation~\eqref{eq:kraus_sampling}, yielding the 
next pure state in the trajectory.
We also refer to quantum trajectories as \emph{noise trajectories}, to emphasize their dependence on stochastic photon-loss events.
We define the final state produced by a quantum trajectory $\tau$ as $\ket{\psi_\tau (\vec{\alpha}, \Theta)}$: although it is unnecessary since it is a pure state, we can also introduce its density matrix as $\rho_\tau(\vec{\alpha}, \Theta) = \ket{\psi_\tau (\vec{\alpha}, \Theta)} \bra{\psi_\tau (\vec{\alpha}, \Theta)}$. By doing this, the expectation over all trajectories of the density matrix of the sampled pure state
results in the density matrix of the prepared mixed state \cite{dalibard_wave-function_1992}: 
\begin{equation}\label{eq:density_matrix_trajectory_wise}
    \rho_\mathrm{final}(\vec{\alpha}, \Theta) = \mathbb{E_\tau[\rho_\tau(\vec{\alpha}, \Theta)]}.
\end{equation}
In this work, we use noise trajectories to define trajectory-wise objective functions and optimize them. When a suitable measurement or heralding mechanism is in place, each single trajectory can then also be associated to an experimental outcome.

\subsection{Risk-Sensitive Optimization}
\label{sec:risk_sensitive_optimization}
In stochastic optimization, objective functions are commonly defined as expectations over the possible outcomes of the underlying random process. This provides a natural criterion for optimizing the average behavior across different realizations.
However, optimizing the expected value may not be sufficient when specific regions of the outcome distribution are of particular interest. For example, in settings such as Reinforcement Learning and Finance, the optimization objective may be designed to penalize unfavorable outcomes more strongly in order to limit potentially large losses \cite{pratt_risk_1964}.
Risk-sensitive optimization provides a way to account for the distribution of the objective values beyond its expectation. Rather than treating all realizations solely through their contribution to the expected value, risk-sensitive objectives can focus on different regions of the distribution. Risk-averse optimization focuses on unfavorable realizations, whereas risk-seeking optimization favors more desirable ones.

An example of risk-sensitive objective is the \emph{entropic risk} measure \cite{howard_risk-sensitive_1972, marthe_efficient_2025}. Let $x(\theta)$ be a random variable representing the value of the quantity of interest under different realizations of the stochastic process, for a given set of model parameters $\theta$. The entropic risk measure is defined as:
\begin{equation}
\label{eq:entropic_risk}
    \mathcal{L}_{\mathrm{ER}}(\theta) = \frac{1}{\beta} \log \mathbb{E}_{x(\theta)}\left[e^{\beta x(\theta)}\right],
\end{equation}
in which the hyperparameter $\beta \in \mathbb{R}$, called \emph{risk coefficient}, controls the risk sensitivity: $\beta > 0$ corresponds to risk-seeking behavior, while $\beta < 0$ leads to risk-averse behavior. In the limit of $\beta \rightarrow 0$, risk-neutral behavior is recovered.

\subsection{Stochastic Computation Graphs}
\label{sec:reinforce_scg}
Consider an objective function defined as an expectation over a discrete random variable $x$ with finite support $\mathcal{X}$:
\begin{equation}
\label{eq:expectation_loss}
    \mathcal{L}(\theta) = \mathbb{E}_{x \sim p_\theta(\cdot)}\left[f(x; \theta)\right].  
\end{equation}
where $p_\theta(x)$ is the probability distribution over $\mathcal{X}$. Both $p_\theta(x)$ and $f(x;\theta)$ depend on the model parameters $\theta$. Consequently, $\mathcal{L}(\theta)$ has a twofold dependence on $\theta$, since changing the parameters affects both the probability $p_\theta(x)$ of sampling each realization $x \in \mathcal{X}$ and the value of $f(x; \theta)$ associated with it.

Stochastic Computation Graphs (SCGs) \cite{schulman_gradient_2016} provide a general framework for differentiating objectives with this kind of twofold dependence, extending the score-function gradient estimator underlying the REINFORCE algorithm \cite{williams_simple_1992} to account also for the explicit parameter dependence of $f$. For Equation~\eqref{eq:expectation_loss}, its unbiased gradient can be written as
\begin{equation}
\label{eq:stochastic_computation_graph}
    \nabla_\theta\, \mathcal{L}(\theta) = \mathbb{E}_{x \sim p_\theta(\cdot)}\left[f(x; \theta)\, \nabla_\theta \log p_\theta(x) + \nabla_\theta\, f(x; \theta)\right],
\end{equation}
in which the first term accounts for the dependence of the sampling distribution on $\theta$, while the second accounts for the explicit dependence of $f$ on $\theta$. 

\section{Risk-Seeking State Preparation with SNAP-Displacement}

In this section, we describe how risk-seeking objective functions, formulated as in Section \ref{sec:risk_sensitive_optimization}, can be applied to the problem of optimizing the parameters of an SD-ansatz. Then, we provide a Monte Carlo procedure that uses quantum trajectories to estimate the gradient needed to optimize such objectives.  

\subsection{Risk-Seeking Objectives on Quantum Trajectories}
\label{sec:losses_on_trajectories}
In a noisy setting, the parameters $\vec{\alpha}, \Theta$ of an SD-ansatz are optimized by maximizing the \emph{average fidelity}
$\hat{F}(\vec{\alpha}, \Theta) = \bra{\psi_{\mathrm{target}}} \rho_{\mathrm{final}}(\vec{\alpha}, \Theta) \ket{\psi_{\mathrm{target}}}$, i.e. the fidelity between the output 
density matrix $\rho_{\mathrm{final}}(\vec{\alpha}, \Theta)$ and the pure target state $\ket{\psi_{\mathrm{target}}}$. The corresponding \emph{average infidelity} is defined as $\hat{I}(\vec{\alpha}, \Theta) = 1-\hat{F}(\vec{\alpha}, \Theta)$. 
\begin{equation}
\label{eq:density_matrix_fidelity}
\begin{aligned}
\mathcal{L}_{\mathrm{avg}}(\vec{\alpha}, \Theta)
=
\bra{\psi_{\mathrm{target}}}
\rho_{\mathrm{final}}(\vec{\alpha}, \Theta)
\ket{\psi_{\mathrm{target}}},
\end{aligned}
\end{equation}
In this work, we propose a different approach, aimed at maximizing the fidelities of only a subset of trajectories. In particular, we aim to increase the probability of obtaining trajectories with very high fidelity, even when this comes at the cost of reducing the average fidelity.
Such a reframing of the problem leads to a risk-seeking formulation for the optimization of SD-ansatzes for state preparation in noisy settings. To do so, first we define the \emph{pure-state fidelity} $F(\vec{\alpha}, \Theta, \tau) = \left|\braket{\psi_\tau (\vec{\alpha}, \Theta) | \psi_{\mathrm{target}}}\right|^2$ as the fidelity between the pure state produced by trajectory $\tau$ and the target state. The pure-state infidelity is then given by $I(\vec{\alpha}, \Theta, \tau) = 1 - F(\vec{\alpha}, \Theta, \tau)$.
We thus define two objective functions by applying the entropic risk measure defined in Equation~\eqref{eq:entropic_risk} to two different quantities of interest. The first uses the pure-state fidelity directly and is referred to as the \emph{fidelity entropic risk}. The second uses the negative logarithm of the pure-state infidelity and is referred to as the \emph{log-infidelity entropic risk}. 
\begin{subequations}
    \begin{center}
        \begin{minipage}{0.4\textwidth}
            \begin{equation}
            \label{eq:sample_entropic_risk_fidelity} 
               \mathcal{L}_F(\vec{\alpha}, \Theta) = \frac{1}{\beta} \log \mathbb{E}_\tau\left[e^{\beta F(\vec{\alpha}, \Theta, \tau)}\right]
            \end{equation}
        \end{minipage}
        \begin{minipage}{0.55\textwidth}
            \begin{equation}
            \label{eq:sample_entropic_risk_log_infidelity} 
                \mathcal{L}_{\log I}(\vec{\alpha}, \Theta) = \frac{1}{\beta} \log \mathbb{E}_\tau\left[e^{-\beta \log(1 - F(\vec{\alpha}, \Theta, \tau))}\right]
            \end{equation}
        \end{minipage}
    \end{center}
\end{subequations}
As explained in Section \ref{sec:risk_sensitive_optimization}, the risk coefficient $\beta > 0$ controls the strength of the risk-seeking behavior: hence, larger values of $\beta$ give relatively more weight to high-fidelity trajectories.
The fidelity entropic risk and the log-infidelity entropic risk objectives differ in the quantity they prioritize. Optimizing the fidelity entropic risk from Equation \eqref{eq:sample_entropic_risk_fidelity} emphasizes improvements in the absolute value of the pure-state fidelity. Optimizing the log-infidelity entropic risk, instead, emphasizes improvements in the order of magnitude of the pure-state infidelity.
Given their definition, both fidelity and log-infidelity entropic risks assign relatively larger values to trajectories with higher pure-state fidelity, with this effect becoming stronger as such a measure approaches $1$.

\subsection{Gradient Computation}
\label{sec:gradient_computation}

Computing the exact values and gradients of the fidelity and log-infidelity entropic risk objectives defined in Equations~\eqref{eq:sample_entropic_risk_fidelity} and~\eqref{eq:sample_entropic_risk_log_infidelity} requires evaluating expectations over all possible quantum trajectories $\tau$ induced by the noise process. Since the number of such trajectories grows combinatorially with the depth of the SD-ansatz, their exact evaluation becomes computationally infeasible.

We therefore estimate the gradients by Monte Carlo sampling of quantum trajectories \cite{lidar_lecture_2020, dum_monte_1992, gardiner_wave-function_1992, abdelhafez_gradient-based_2019, cheng_efficient_2023}. Assuming Markovianity of the quantum evolution caused by photon loss, as previous works argued \cite{wiseman_quantum_1996, gardiner_wave-function_1992}, quantum trajectories are sampled by evolving the pure state of the qudit stochastically according to the probabilities $p_j$ of applying Kraus operator $K_j$ after each SD-block, as described in Section \ref{sec:noise_model}.

The resulting optimization problem has the structure introduced in Section~\ref{sec:reinforce_scg}: the SD-ansatz parameters $\vec{\alpha},\Theta$ affect both the probability distribution over noise trajectories and the objective value associated with each sampled trajectory.
We thus use the SCG expression in Equation~\eqref{eq:stochastic_computation_graph} to construct a Monte Carlo gradient estimator for the fidelity and log-infidelity entropic risk objectives. 
While the SCG estimator is unbiased for expectations, the logarithm outside the expectation generally introduces finite-sample bias. We nevertheless use the resulting estimator as an optimization objective.

\section{Experimental Protocol}
\label{sec:experimental_setup}

In our experiments, we consider one qudit of $d=8$ levels in the Fock space and a SD-ansatz with $B=4$ SD-blocks, as defined in Equation \eqref{eq:snap_displacement_protocol}.
To train the parameters of this ansatz, we use the objective functions that we have introduced in Section \ref{sec:losses_on_trajectories}, defined over trajectories sampled according to the noise model of photon-counting Kraus operators. Throughout this work, we set $T_1=1ms$, a regime that has been shown in previous numerical studies to constitute a challenging state preparation setting \cite{dacrema_sparse_2026}. We simulate on a classical computer how the pure state of the qudit evolves, as described in Section \ref{sec:noise_model}, by applying the unitaries of the S and D gates within each SD-block, interleaving the SD-blocks with applications of our noise model throughout the whole SD-ansatz. This is consistent with what is done in \cite{dacrema_sparse_2026}: just like they do, we also include in the simulation $64$ additional bumper states, to take into account the leakage described in Section~\ref{sec:snap_displacement}.
For the gradient estimation, we collect $10$ quantum trajectories at each training step through Monte Carlo sampling, as explained in Section \ref{sec:gradient_computation}.
The qudit state is initialized as the Fock state $\ket{0}$, and we aim to prepare as a target a \emph{Haar random} pure state, which is chosen from the columns of a Haar random unitary generated according to \cite{mezzadri_how_2007}.

To have statistically robust results, we run 100 independent optimizations of the parameters for each SD-ansatz, starting from different random initializations, and we call these \emph{repetitions}.
We train each repetition for $30000$ steps, using the Adam optimizer with learning rate $10^{-2}$, momentum parameters $(0.9, 0.999)$ and no regularization.
We define the \emph{$\epsilon$-probability} as the probability of sampling a noise trajectory with pure-state infidelity below a threshold $\epsilon$, which we call \emph{infidelity threshold}. 
For each repetition, we estimate the $\epsilon$-probability by sampling 10000 noise trajectories and computing the fraction of trajectories with final pure-state infidelity below $\epsilon$.
We consider infidelity thresholds ranging from $\epsilon=10^{-1}$ to $\epsilon=10^{-7}$.
Due to the stochasticity of the optimization process, though, only some repetitions yield an SD-ansatz with a non-zero $\epsilon$-probability. We refer to them as \emph{successful repetitions}, and define the \emph{success rate} (SR) as their percentage.
For each objective function and value of $\beta$, we define the \emph{minimum infidelity threshold} as the lowest infidelity threshold reached with a non-zero SR.
The mean and standard deviation of the $\epsilon$-probability are computed over successful repetitions only, and are thus conditional on successful optimization.
We argue that this is reasonable, as the SR and the $\epsilon$-probability capture two distinct aspects of the optimization process. Specifically, the first reflects the reliability of such a process, while the second quantifies the probability of reaching a desired infidelity threshold using the parameters corresponding to any successful repetition: in practice, one would use only such parameters. 
We train the SD-ansatz using the fidelity and log-infidelity entropic risk objectives for multiple values of the risk coefficient $\beta$. For each objective and value of $\beta$, we compute the $\epsilon$-probabilities and SR and we analyze their trade-off with respect to the infidelity thresholds $\epsilon$.
For comparison, we use as a baseline the $\epsilon$-probabilities and success rates obtained by training the SD-ansatz using the average fidelity objective.

\section{Results \& Discussion}

\begin{table}[ht]
\centering
\setlength{\tabcolsep}{3pt}
\renewcommand{\arraystretch}{1.15}

\resizebox{\textwidth}{!}{
\begin{tabular}{
    >{\centering\arraybackslash}p{2.1cm}|
    >{\centering\arraybackslash}p{1.1cm}|
    >{\centering\arraybackslash}p{2.05cm}|
    *{6}{>{\centering\arraybackslash}p{1.4cm}|}
    >{\centering\arraybackslash}p{1.4cm}
}
\hline

\multirow[c]{2}{2.1cm}{\centering Objective function}
&
\multirow[c]{2}{1.0cm}{\centering $\beta$}
&
\multirow[c]{2}{2.05cm}{\centering Average fidelity}
&
\multicolumn{7}{c}{$\epsilon$-probability for an infidelity $\epsilon$ of:}
\\

\cline{4-10}

&
&
&
$10^{-1}$
&
$10^{-2}$
&
$10^{-3}$
&
$10^{-4}$
&
$10^{-5}$
&
$10^{-6}$
&
$10^{-7}$
\\

\hline

\centering Average\\fidelity
&
\textbf{N/A}
&
0.823921 $\pm$ 0.035875
&
\underline{0.793} $\pm$ 0.041 (SR=86\%)
&
\textbf{\underline{0.772} $\pm$ 0.026 (SR=5\%)}
&
0.0 $\pm$ 0.0 (SR=0\%)
&
0.0 $\pm$ 0.0 (SR=0\%)
&
0.0 $\pm$ 0.0 (SR=0\%)
&
0.0 $\pm$ 0.0 (SR=0\%)
&
0.0 $\pm$ 0.0 (SR=0\%)
\\

\hline

\multirow[c]{5}{2.1cm}{
    \parbox[c]{2.0cm}{
        \centering
        Fidelity\\
        entropic\\
        risk
    }
}
&
\textbf{10}
&
0.795007 $\pm$ 0.079314
&
{0.766} $\pm$ 0.085
({SR=97\%})
&
{0.766} $\pm$ 0.049
({SR=86\%})
&
\textbf{\underline{0.776} $\pm$ 0.034 (SR=32\%)}
&
0.0 $\pm$ 0.0 (SR=0\%)
&
0.0 $\pm$ 0.0 (SR=0\%)
&
0.0 $\pm$ 0.0 (SR=0\%)
&
0.0 $\pm$ 0.0 (SR=0\%)
\\

\cline{2-10}

&
\textbf{100}
&
0.728762 $\pm$ 0.139118
&
0.658 $\pm$ 0.196 (SR=95\%)
&
0.726 $\pm$ 0.095 (SR=77\%)
&
0.745 $\pm$ 0.051 (SR=68\%)
&
\textbf{\underline{0.760} $\pm$ 0.039 (SR=39\%)}
&
0.0 $\pm$ 0.0 (SR=0\%)
&
0.0 $\pm$ 0.0 (SR=0\%)
&
0.0 $\pm$ 0.0 (SR=0\%)
\\

\cline{2-10}

&
\textbf{1000}
&
0.698329 $\pm$ 0.144692
&
0.602 $\pm$ 0.204
({SR=97\%})
&
0.682 $\pm$ 0.082 (SR=79\%)
&
0.694 $\pm$ 0.042
({SR=74\%})
&
0.708 $\pm$ 0.028
({SR=61\%})
&
\textbf{\underline{0.713} $\pm$ 0.027 (SR=24\%)}
&
0.0 $\pm$ 0.0 (SR=0\%)
&
0.0 $\pm$ 0.0 (SR=0\%)
\\

\cline{2-10}

&
\textbf{10000}
&
0.687215 $\pm$ 0.134888
&
0.590 $\pm$ 0.178
({SR=97\%})
&
0.638 $\pm$ 0.117 (SR=84\%)
&
0.671 $\pm$ 0.031 (SR=67\%)
&
0.676 $\pm$ 0.030 (SR=58\%)
&
0.681 $\pm$ 0.024
({SR=38\%})
&
\textbf{\underline{0.690} $\pm$ 0.013 (SR=5\%)}
&
0.0 $\pm$ 0.0 (SR=0\%)
\\

\cline{2-10}

&
\textbf{100000}
&
0.663848 $\pm$ 0.156595
&
0.534 $\pm$ 0.220
({SR=97\%})
&
0.625 $\pm$ 0.128 (SR=76\%)
&
0.659 $\pm$ 0.027 (SR=64\%)
&
0.661 $\pm$ 0.027 (SR=56\%)
&
0.672 $\pm$ 0.021 (SR=36\%)
&
\textbf{0.673 $\pm$ 0.008
({SR=7\%})}
&
0.0 $\pm$ 0.0 (SR=0\%)
\\

\hline

\multirow[c]{6}{2.1cm}{
    \parbox[c]{2.0cm}{
        \centering
        Log-infidelity\\
        entropic\\
        risk
    }%
}
&
\textbf{0.01}
&
0.762516 $\pm$ 0.055348
&
{0.704} $\pm$ 0.037 (SR=66\%)
&
{0.695} $\pm$ 0.027 (SR=49\%)
&
{0.694} $\pm$ 0.026 (SR=43\%)
&
{0.699} $\pm$ 0.025 (SR=31\%)
&
\textbf{{0.699} $\pm$ 0.019 (SR=20\%)}
&
0.0 $\pm$ 0.0 (SR=0\%)
&
0.0 $\pm$ 0.0 (SR=0\%)
\\

\cline{2-10}

&
\textbf{0.1}
&
0.775744 $\pm$ 0.032523
&
0.684 $\pm$ 0.041 (SR=92\%)
&
0.676 $\pm$ 0.032 (SR=78\%)
&
0.675 $\pm$ 0.031 (SR=74\%)
&
0.676 $\pm$ 0.029 (SR=63\%)
&
\textbf{0.677 $\pm$ 0.026 (SR=38\%)}
&
0.0 $\pm$ 0.0 (SR=0\%)
&
0.0 $\pm$ 0.0 (SR=0\%)
\\

\cline{2-10}

&
\textbf{1.0}
&
0.748327 $\pm$ 0.032474
&
0.663 $\pm$ 0.036 (SR=97\%)
&
0.663 $\pm$ 0.036 (SR=94\%)
&
0.666 $\pm$ 0.035
(\underline{SR=87\%})
&
0.667 $\pm$ 0.032
(\underline{SR=80\%})
&
0.668 $\pm$ 0.034 (SR=47\%)
&
\textbf{0.680 $\pm$ 0.032
(\underline{SR=12\%})}
&
0.0 $\pm$ 0.0 (SR=0\%)
\\

\cline{2-10}

&
\textbf{2.0}
&
0.750230 $\pm$ 0.032208
&
0.663 $\pm$ 0.036
(\underline{SR=100\%})
&
0.664 $\pm$ 0.034 (SR=93\%)
&
0.667 $\pm$ 0.031 (SR=85\%)
&
0.672 $\pm$ 0.025 (SR=74\%)
&
0.679 $\pm$ 0.022 (SR=48\%)
&
\textbf{0.682 $\pm$ 0.017 (SR=10\%)}
&
0.0 $\pm$ 0.0 (SR=0\%)
\\

\cline{2-10}

&
\textbf{5.0}
&
0.742331 $\pm$ 0.041070
&
0.659 $\pm$ 0.038 (SR=99\%)
&
0.661 $\pm$ 0.038 (SR=94\%)
&
0.666 $\pm$ 0.032 (SR=85\%)
&
0.670 $\pm$ 0.027 (SR=75\%)
&
0.673 $\pm$ 0.024 (SR=42\%)
&
\textbf{0.685 $\pm$ 0.020 (SR=7\%)}
&
0.0 $\pm$ 0.0 (SR=0\%)
\\

\cline{2-10}

&
\textbf{10.0}
&
0.719670 $\pm$ 0.096950
&
0.615 $\pm$ 0.143 (SR=99\%)
&
0.618 $\pm$ 0.137
(\underline{SR=95\%})
&
0.622 $\pm$ 0.142
(\underline{SR=87\%})
&
0.645 $\pm$ 0.098 (SR=75\%)
&
0.652 $\pm$ 0.085
(\underline{SR=51\%})
&
\textbf{0.681 $\pm$ 0.018 (SR=6\%)}
&
0.0 $\pm$ 0.0 (SR=0\%)
\\

\hline
\end{tabular}
}
\caption{$\epsilon$-probabilities and success rates (SR) computed from the SD-ansatzes trained on the objective functions listed in Section~\ref{sec:losses_on_trajectories}, for different values of the risk coefficient $\beta$, if present.
Each row reports the results of training the parameters of 100 repetitions of an SD-ansatz.
For each objective function, the column "Average fidelity" reports the mean average fidelity across all 100 repetitions, with its standard deviation.
The group of columns under the "$\epsilon$-probability for an infidelity $\epsilon$ of:" label reports the evaluation using Monte Carlo sampling of noise trajectories, as explained in Section \ref{sec:experimental_setup}. 
The mean probability estimates are computed using only the successful repetitions, which are also the same repetitions indicated by the SR. For each objective function and risk coefficient value, we highlight in \textbf{bold} the cell related to the lowest infidelity achieved by them. For each infidelity threshold instead, we \underline{underline} the values related either to the highest probability estimate or the highest SR.}
\label{tab:entropic_risk_objectives}
\end{table}
The goal of this study is to assess whether, in noisy settings, an SD-ansatz optimized with risk-seeking objectives can prepare states with pure-state infidelity lower than that achieved when using the average fidelity. 
We report the full experimental results in Table~\ref{tab:entropic_risk_objectives}. First, we give an overview of the results for each objective function. Then, we analyze the trade-off between average fidelity and pure-state infidelity. Next, we examine the trade-off between the infidelity threshold $\epsilon$, the corresponding $\epsilon$-probability, and the success rate (SR). Finally, we discuss possible interpretations of the observed results.

\subsection{Overview of the Objective Functions}
First of all, we look at the average fidelity objective function. The SD-ansatzes prepared using this objective yield the best average fidelity value, $0.823921$, across all experiments, when compared to both risk-seeking objectives and for all the values of their risk coefficient $\beta$. 
However, the parameters of the trained SD-ansatzes start to behave poorly very quickly: not only $10^{-2}$ is its minimum infidelity threshold, but also only 5\% of the repetitions reach that.
Yet, for the successful repetitions, the $\epsilon$-probabilities for these relatively large infidelities are all rather high, between $0.77$ and $0.80$.

Optimizing on the fidelity entropic risk objective, instead, leads to a degradation in the computed average fidelity of the prepared ensemble. $\beta = 10$ achieves an average fidelity of $0.795007$ which is only slightly smaller than that achieved by the previous objective; however, the average fidelity steadily drops as the value of the risk coefficient $\beta$ increases, all the way down to $0.663848$ for $\beta = 100000$, and we can see that its standard deviation also increases, indicating diversity in this metric across the single repetitions. 
At the same time, this objective is able to push SD-ansatzes towards yielding pure states with infidelities as low as $10^{-6}$. For reaching such a threshold, though, the risk coefficient $\beta$ needs to be quite large, i.e. $\beta = 10000$, and the SR is still only $5\%$, which is very low. Moreover, the SR for the $10^{-6}$ infidelity threshold saturates beyond a certain value of $\beta$, and does not further improve: even $\beta = 100000$ cannot raise the SR beyond $7\%$, nor reach $\epsilon = 10^{-7}$. Instead, the SR increases considerably when aiming for larger thresholds, for example reaching $61\%$ for $\epsilon = 10^{-4}$ and $\beta = 1000$. 
Another thing to note is that the $\epsilon$-probabilities for the largest infidelity thresholds also drop when the risk coefficient $\beta$ grows, reaching the lowest value of $0.534$ when $\epsilon = 10^{-1}$ and $\beta = 100000$. On the contrary, smaller values of $\beta$, for example $\beta = 10$, lead to $\epsilon$-probabilities comparable to those estimated for the average fidelity objective. 
Also, as the desired infidelity threshold $\epsilon$ gets lower, the $\epsilon$-probabilities instead increase, while their standard deviations decrease: for example, for $\beta = 100000$, the $\epsilon$-probabilities improve from $0.534 \pm 0.220$ to $0.673 \pm 0.008$. In fact, when considering $10^{-6} \leq \epsilon \leq 10^{-3}$ the largest $\epsilon$-probabilities
are achieved by the fidelity entropic risk objective and are \underline{underlined} in Table \ref{tab:entropic_risk_objectives} as such.

Last, we analyze the results for the log-infidelity entropic risk objective function. As we can see in Table~\ref{tab:entropic_risk_objectives}, its behavior is largely consistent with that of the fidelity entropic risk. However, compared to it, as the value of the risk coefficient $\beta$ gets larger, the decrease in the average fidelity is smaller and slower, going only from $0.762516$ for $\beta = 0.01$ to $0.719670$ for $\beta = 10.0$. Moreover, the infidelity threshold $\epsilon = 10^{-6}$ is reached already with $\beta = 1.0$, much smaller than that required by the fidelity entropic risk, even though the same saturation phenomenon prevents achieving a threshold of $10^{-7}$. Furthermore, the SR of the training process for the log-infidelity entropic risk is \underline{underlined} in Table \ref{tab:entropic_risk_objectives} as the largest among all the objective functions for \emph{all} the infidelity thresholds.
For example, even infidelity thresholds as low as $10^{-4}$ and $10^{-5}$ have respectively the highest SR of $80\%$ for $\beta = 1.0$, and $51\%$ for $\beta = 10.0$ . 
However, this comes at the cost of slightly smaller $\epsilon$-probabilities. In fact, for all the infidelity thresholds, the best $\epsilon$-probabilities have values between $0.68$ and $0.7$, and \emph{never} exceed the best ones achieved by the other objectives. Still, the trends in the $\epsilon$-probabilities of the log-infidelity entropic risk are the same, but less pronounced, as those observed for the fidelity entropic risk for $\beta$ increasing or for considering progressively lower infidelity thresholds. As an example, for the threshold $\epsilon = 10^{-1}$ the $\epsilon$-probabilities drop from $0.704$ to $0.615$ as $\beta$ increases. Instead, choosing a constant $\beta = 10.0$ leads to the $\epsilon$-probabilities increasing from $0.615$ to $0.681$ as the threshold $\epsilon$ gets smaller.

\subsection{Trade-Off between Average Fidelity and Pure-State Fidelity}

With risk-seeking objective functions, we have observed a decrease of \textbf{up to four orders of magnitude} in the minimum infidelity threshold achieved. However, at the same time, the average fidelity also gets smaller in general, compared to that measured using the average fidelity itself as objective, with its standard deviation increasing as well.
This degradation in the average fidelity happens in tandem with the improvement in the minimum infidelity threshold as the risk coefficient $\beta$ gets larger, and is much faster for the fidelity entropic risk. Nonetheless, both the decrease in the average fidelity and the increase in its standard deviation are less impactful, but still present, for the log-infidelity entropic risk. 
Our results empirically confirm that, in the considered experimental setting, risk-seeking optimization exposes a trade-off between average fidelity and pure-state fidelity, enabling higher pure-state fidelities to be reached with a certain probability at the cost of reducing the average fidelity.
Furthermore, such a trade-off between the average fidelity and the minimum infidelity threshold can be controlled through the risk coefficient $\beta$.
These results are consistent with how risk-sensitive optimization behaves in general, as already explained in Section \ref{sec:risk_sensitive_optimization}.

\subsection{Trade-Off between Infidelity Threshold, $\epsilon$-Probability and Success Rate} 

The two entropic risk objective functions expose different operating regimes. The fidelity entropic risk generally yields SD-ansatzes with higher $\epsilon$-probabilities for any given infidelity threshold $\epsilon$, whereas the log-infidelity entropic risk consistently achieves higher success rates, resulting more often in SD-ansatzes that reach the desired threshold. 
In general, for a fixed infidelity threshold $\epsilon$, the $\epsilon$-probability tends to decrease as $\beta$ increases. At the same time, increasing $\beta$ generally enables lower minimum infidelity thresholds.
The reduction of $\epsilon$-probabilities is more pronounced for the fidelity entropic risk. 
For both risk-seeking objectives, the SR initially increases with $\beta$ for a fixed infidelity threshold, but eventually saturates or even decreases.
These observations identify a trade-off between reaching lower minimum infidelity thresholds, maintaining large enough $\epsilon$-probabilities and achieving a consistent SR of the training process. 
Such a trade-off can be controlled by modifying the risk coefficient $\beta$.

\subsection{Discussion}

Overall, optimizing on the entropic risk objectives yields remarkably lower infidelities than training on the average fidelity objective does, with an improvement as large as four orders of magnitude in the lowest infidelity reached. 
One possible interpretation is that optimizing the shared ansatz parameters for different noise trajectories may lead to conflicting effects on their fidelities, preventing them from being simultaneously maximized. While the average fidelity would favor a compromise across noise trajectories, the entropic risk objectives may instead prioritize a subset of trajectories whose fidelities can be further increased at the expense of the others.
This aligns with our observation that the average fidelity objective does not push the pure-state infidelity towards the lowest thresholds.

A second behavior can also be observed when evaluating progressively smaller infidelity thresholds for each value of $\beta$.
Indeed, although lower thresholds correspond to more challenging tasks, the $\epsilon$-probabilities often remain roughly constant or even increase, while their standard deviation decreases.
While this might seem surprising at first, it should instead be interpreted together with the decreasing SR, since the $\epsilon$-probabilities are computed only over successful repetitions. Thus, the observed behavior suggests that the repetitions that remain successful at lower infidelity thresholds assign a large probability mass to states with very low pure-state infidelity. However, this behavior may also be an artifact due to the progressively fewer successful repetitions, and should be further investigated.

For any chosen infidelity threshold, we also observed that the $\epsilon$-probability generally decreases as $\beta$ increases, while its standard deviation tends to increase. This means that the $\epsilon$-probability becomes more variable across successful repetitions. 
Moreover, especially for lower infidelity thresholds, increasing $\beta$ beyond a certain value may reduce the SR without providing a corresponding improvement in the $\epsilon$-probability or in the lowest achievable pure-infidelity threshold, which eventually saturates. Therefore, larger risk coefficients do not necessarily lead to more favorable operating points, and $\beta$ should be carefully selected to match the desired trade-off between the target infidelity, $\epsilon$-probability, and SR. 
At the same time, one should select the objective function depending on the application to balance the $\epsilon$-probability against the SR of the training. If maximizing the $\epsilon$-probability of the prepared ensemble for a target infidelity $\epsilon$ is the primary objective, the fidelity entropic risk may be preferable. Instead, the log-infidelity entropic risk is more reliable in training an SD-ansatz that reaches the target infidelity. 

\section{Conclusions \& Future Directions}

In this work, we introduced a risk-seeking optimization framework for noisy state preparation in bosonic quantum systems. 
By optimizing trajectory-wise objectives through Monte Carlo sampling, our approach provides control over the distribution of pure-state fidelities.
Furthermore, our results show that risk-seeking optimization can reduce the lowest achievable pure-state infidelity by \textbf{up to four orders of magnitude} compared to standard optimization of the average fidelity in the considered experimental setting.
More importantly, we show that the proposed framework provides control over the trade-off between the pure-infidelity threshold $\epsilon$, the $\epsilon$-probability 
and the success rate of the optimization process. 
In particular, the optimization objective and its risk coefficient $\beta$ allow the operating point of the state-preparation procedure to be tailored to the requirements of a specific application.
Future work should investigate whether the observed behaviors generalize to different target states, larger Fock-space dimensions, deeper SD-ansatzes and different noise models. 
Finally, comparing the quantum states prepared by our framework with those obtained by optimizing the average fidelity may reveal which state properties are favored by trajectory-wise optimization, guiding the design of alternative risk-sensitive objectives.

\begin{acknowledgments}
    We acknowledge ISCRA for awarding this project access to the LEONARDO supercomputer, owned by the EuroHPC Joint Undertaking, hosted by CINECA (Italy).
\end{acknowledgments}

\section*{Declaration on Generative AI}
  
 During the preparation of this work, the authors used GPT-5.6 Sol, GPT-5.6 Luna and Grammarly in order to paraphrase and reword, as well as grammar and spelling check. 
 After using these tools/services, the authors reviewed and edited the content as needed and take full responsibility for the publication’s content. 

\bibliography{references}


\end{document}